\documentclass[a4paper, conference]{IEEEtran}
\IEEEoverridecommandlockouts
\usepackage{cite}
\usepackage{graphicx}

\def\BibTeX{{\rm B\kern-.05em{\sc i\kern-.025em b}\kern-.08em
        T\kern-.1667em\lower.7ex\hbox{E}\kern-.125emX}}
\usepackage{amsmath}
\usepackage{mathtools}
\usepackage{algpseudocode}
\usepackage{algorithm}
\usepackage{algorithmicx}
\usepackage{textcomp}
\usepackage{filecontents}
\usepackage{microtype}
\usepackage{float}
\usepackage{adjustbox}
\usepackage{booktabs,makecell,tabularx}
\usepackage{url}
\usepackage{booktabs}
\usepackage{subfigure}

\usepackage{amssymb}

\newcolumntype{C}[1]{>{\centering\arraybackslash}p{#1}}
\newcolumntype{L}{>{\raggedright\arraybackslash}X}
\usepackage{siunitx}
\usepackage[utf8]{inputenc}
\renewcommand{\thetable}{\arabic{table}}
\usepackage{etoolbox}
\usepackage{xparse}
\makeatletter

    \makeatletter
    \newcommand{\linebreakand}{%
      \end{@IEEEauthorhalign}
      \hfill\mbox{}\par
      \mbox{}\hfill\begin{@IEEEauthorhalign}
    }
    \makeatother

\begin{document}




\title{
Fighting Game Commentator with Pitch and Loudness Adjustment Utilizing Highlight Cues

















}
    \author{
\IEEEauthorblockN{Junjie H. Xu\IEEEauthorrefmark{1}, Zhou Fang\IEEEauthorrefmark{1}, Qihang Chen\IEEEauthorrefmark{2}, Satoru Ohno\IEEEauthorrefmark{1}, Pujana Paliyawan\IEEEauthorrefmark{3}}
\IEEEauthorblockA{
\IEEEauthorrefmark{1}Graduate School of Comprehensive Human Sciences, University of Tsukuba, Japan\\
\IEEEauthorrefmark{2}Department of Sociology, Doshisha University, Japan\\
\IEEEauthorrefmark{3}Research Organization of Science and Technology, Ritsumeikan University, Japan\\
s2021705@s.tsukuba.ac.jp}
}


    \maketitle

\begin{abstract}

This paper presents a commentator for providing real-time game commentary in a fighting game. The commentary takes into account highlight cues, obtained by analyzing scenes during gameplay, as input to adjust the pitch and loudness of commentary to be spoken by using a Text-to-Speech (TTS) technology. We investigate different designs for pitch and loudness adjustment. The proposed AI consists of two parts: a dynamic adjuster for controlling pitch and loudness of the TTS and a real-time game commentary generator. We conduct a pilot study on a fighting game, and our result shows that by adjusting the loudness significantly according to the level of game highlight, the entertainment of the gameplay can be enhanced.


\end{abstract}


\section{Introduction}

Watching video game live-streaming via platforms such as Twitch and YouTube has snowballed in popularity for a decade, and it has become a new kind of entertainment\cite{2012pro} with a considerable market value\cite{johnson2019impacts}. The game commentary can keep game live-streaming audiences entertained and informed~\cite{whyesports}. However to employ human commentator is costly, and therefore, the demand for non-human or AI commentators has been surfaced and increasingly gained interest from researchers~\cite{Esports}. As game commentary is a kind of expressive speech, synthesizing game commentary requires not only synthesizing realistic speech using text input from game scenes but also adjusting the phonetic variance that expresses the emotional information based on the context~\cite{emotionalEffects}. 

This paper presents the design of a game commentary system for providing game commentary during gameplay. We proposed a phonetic adjuster that adjusts the pitch and the loudness of speech based on highlight cues \cite{ishii2019fighting}. Investigation on five different designs for pitch and loudness adjustment was conducted. The study uses a fighting game platform called FightingICE\footnote{https://www.ice.ci.ritsumei.ac.jp/$\sim$ftgaic/index-2.html}, which is commonly used for game AI research.

The main contributions of this work are as follows:
\begin{enumerate}
    \item the design of the proposed method that dynamically adjusts the pitch and volume of a given commentary based on highlight cues, 
    \item the description of the real-time commentary generation system in use, and 
    \item the study on the effects of pitch and/or volume adjustment to the spectator's entertainment
\end{enumerate}

\section{Existing Work}

In traditional sports, some research investigated commentaries by human commentators from the perspective of their phonetic variation~\cite{behind}. Commercial games in recent decades, such as NBA 2K series and FIFA series, commentaries by human commentators were pre-recorded then replayed during the gameplay. Therefore, the demand for building intelligent live commentary generating systems for video game live-streaming, expected to bring higher productivity at lower costs than human commentators, has surfaced and gained much interest by researchers\cite{ppg, guzdial2019play, kameko2015learning, Esports}. The recent advancement on neural models accelerated the development of synthesizing natural human speech~\cite{wavenet}, enabling the possibility to synthesize expressive speech such as commentary. Tanaka et al.~\cite{caption} focused on providing textual information of commentary for esports videos, Shah et al.~\cite{style} built a live commentary generating system for video game live-streaming, Jemneanbun et al.~\cite{tj} integrated the commentary texts with rap music. However, controlling prosody such as pitch and loudness for commentary to respond to real-time context (highlight of the game scene in this work) in videos or games has not been well-investigated. Table \ref{tab:existingwork} shows the comparison of this work to existing works.

\begin{table}[tb]
    \caption{Comparison of study conducted in this work to existing works over 5 dimentions: (A) Real-time, (B) Artificial Intelligence Agent, (C) Speech Synthesis, (D) Game Commentary, (E) Context-Oriented Phonetic Adjustment}
    \label{tab:existingwork}
      \hbox to\hsize{\hfil
  \begin{tabular}{lccccc}
    \toprule
    Comparison works & A & B & C & D & E\\
    \midrule
    Ishii et al.~\cite{ishii2019fighting}& \checkmark & \checkmark & - & - & -\\ 
    Tanaka et al.~\cite{caption} & - & - & - & \checkmark & -\\ 
    Shah et al.~\cite{style} & \checkmark & - & \checkmark & \checkmark & -\\ 
    Jemneanbun et al.~\cite{tj} & \checkmark & \checkmark & \checkmark & \checkmark & - \\ 
    \textbf{Ours} & \checkmark & \checkmark & \checkmark & \checkmark & \checkmark\\
    \bottomrule
    \hline
  \end{tabular}\hfil}
\end{table}

\section{Proposed AI Commentator System}
\begin{figure}[tb]
  \centering
  \includegraphics[width=\linewidth]{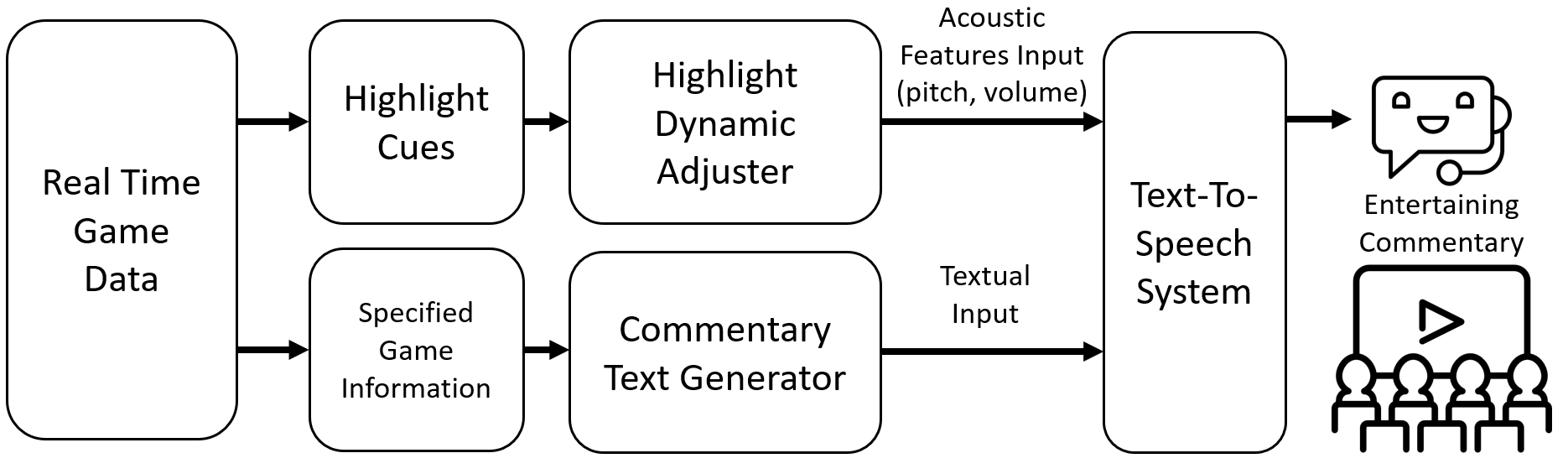}
  \caption{Overview of Proposed Commentary System}
  \label{fig:system}
\end{figure}

The proposed system (demo available\footnote{https://tinyurl.com/2jacub2t}) consists of two modules as shown in Fig. \ref{fig:system}: the highlight dynamic phonetic adjuster and a commentary text generator based on templates. Each commentary is generated by four steps as follows:
\begin{description}
\item[Step 1]: accesses game data in real-time and if TTS is not playing, go to Step 2; otherwise, repeat Step 1. 
\item[Step 2]: calculates both pitch and volume values according to the proposed phonetic adjuster and at the same time generates a commentary text.
\item[Step 3]: sends the resulting values of the two phonetic parameters and the generated commentary text to TTS via JSON HTTP requests.
\item[Step 4]: passes the audio output from TTS to the spectators.
\end{description}
The main hypothesis behind the AI commentator system, in particular, the proposed phonetic adjuster, is that excitement in a fight between two players in the fighting game can be numerically derived from highlight cues and the resulting value can then be used to increase or decrease the values of relevant phonetic parameters. 
\begin{table*}[htbp]
    \caption{Examples of using system generating commentaries using proposed method}
    \label{tab:exp}
       \hbox to\hsize{\hfil
  \begin{tabular}{l c c c c c c c c c c}
    \toprule
     & \multicolumn{10}{c}{Scenes}
    \\

     & \multicolumn{5}{c}
    {
    \begin{minipage}{.4\textwidth}
      \includegraphics[width=\linewidth]{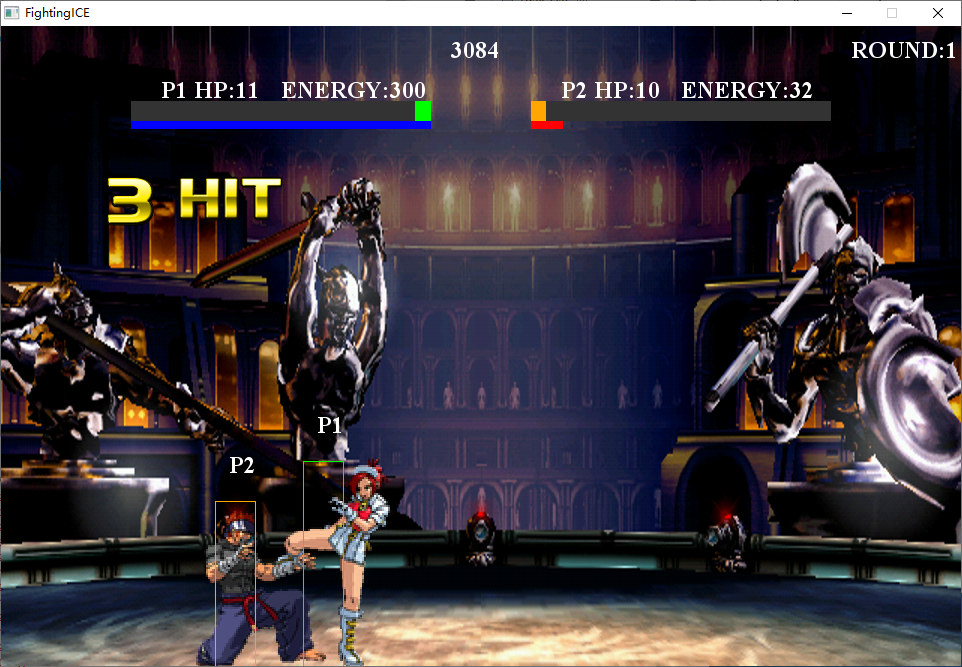}
    \end{minipage} 
    } 
    & \multicolumn{5}{c}
    {
        \begin{minipage}{.4\textwidth}
      \includegraphics[width=\linewidth]{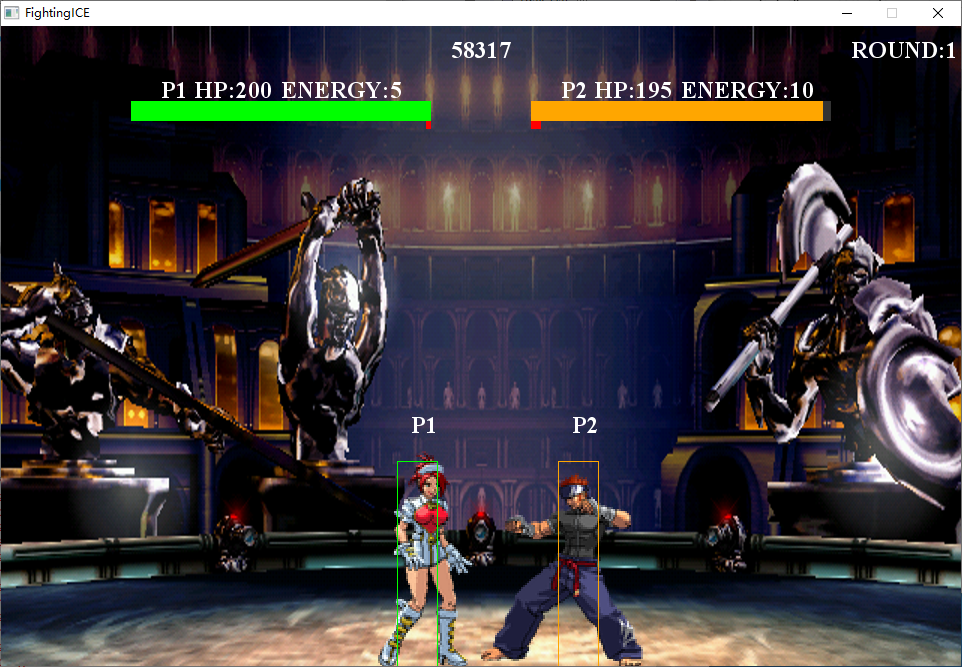}
    \end{minipage}    
    } 
    \\
    \\
     Scenes & \multicolumn{5}{c}{Example scene with high value} & \multicolumn{5}{c}{Example scene with low value}  
    \\
    
     Texts & \multicolumn{5}{c}{Garnet is so powerful releasing Heavy Kick} & \multicolumn{5}{c}{They try to predict each other!}  
     \\
           & \multicolumn{5}{c}{that Zen should be very careful!} & \multicolumn{5}{c}{}  

    \\
    \\
    
    $Score$ & \multicolumn{5}{c}{0.9973} & \multicolumn{5}{c}{0.0402} 
    \\
 
    $Action$ & \multicolumn{5}{c}{0} & \multicolumn{5}{c}{0}  
    \\

    $Distance$ & \multicolumn{5}{c}{0.8743} & \multicolumn{5}{c}{0.6646}  
    \\

      \textbf{Highlight Values} & \multicolumn{5}{c}{\textbf{0.6239}} & \multicolumn{5}{c}{\textbf{0.2349}}  
    \\ 
    \\ 
   
     
     
     
    
        \midrule
   
     & Design 1 & Design 2 & Design 3 & Design 4 & Design 5 & Design 1 & Design 2 & Design 3 & Design 4 & Design 5  \\
    \midrule
   When highlight Value $\uparrow$ & $-$ &  Volume $\uparrow$ & Volume $\downarrow$ & Pitch $\uparrow$ & Pitch $\downarrow$ & $-$ &  Volume $\uparrow$ & Volume $\downarrow$ & Pitch $\uparrow$ & Pitch $\downarrow$  \\
    Input Value of Volume & \textit{Default} & \textbf{0.9909} & \textbf{-0.9909} & \textit{Default} & \textit{Default} & \textit{Default} & \textbf{-2.1205} & \textbf{2.1205} & \textit{Default} & \textit{Default}  \\
    Input Value of Pitch & \textit{Default} & \textit{Default} & \textit{Default} & \textbf{6.4772} & \textbf{1.5228} & \textit{Default} & \textit{Default} & \textit{Default} & \textbf{-1.3015} & \textbf{9.3014}\\
    
    \bottomrule
    \hline
    \end{tabular}\hfil}
\end{table*}

\begin{table}[tb] 
  \caption{$RankAct$ List in descending damage values~\cite{ishii2019fighting}} 
  \label{tab:rankact}
  \hbox to\hsize{\hfil
  \begin{tabular}{c c c}
    \toprule
    $RankAct$ & Skill Content & $Rank$ \\
    \midrule
    STAND\underline{\hspace{0.5em}}D\underline{\hspace{0.5em}}DF\underline{\hspace{0.5em}}FC & Special Skill & 1 \\
    STAND\underline{\hspace{0.5em}}F\underline{\hspace{0.5em}}D\underline{\hspace{0.5em}}DFB & Strong Upper & 2 \\
    STAND\underline{\hspace{0.5em}}D\underline{\hspace{0.5em}}DB\underline{\hspace{0.5em}}BB & Sliding Kick & 3 \\
    STAND\underline{\hspace{0.5em}}D\underline{\hspace{0.5em}}DF\underline{\hspace{0.5em}}FB & Shoot Strong Projectile Forward & 4\\
    \bottomrule
    \hline
  \end{tabular}\hfil}
\end{table}

\subsection{Highlight Cues}

Three highlight cues, ``Score Transition'' ($Score$), ``Action'' ($Action$), and ``Distance'' ($Distance$), are used in highlight evaluation. Namely, (1) $Score$ promotes score transition near the end of a fight round by multiplying the elapsed time of the current round and the predicted damage value on the opponent's Hit Point (HP) of a selected action, min-max normalization is applied to both of them, (2) follows Ishii et al.~\cite{ishii2019fighting}, $Action$ prioritizes actions that considered more entertaining, (3) $Distance$ giving prominence to the close-distance fight using the value of dividing the current distance between P1 and P2 by the maximum distance in the current round. $Highlight$ is the summation of the normalized values of $Score$, $Action$ and $Distance$.

\subsubsection{Score}
As with Ishii et al.~\cite{ishii2019fighting}, this cue promotes score transition near the end of a round. It is implemented as the equation follows:

\begin{equation}
\begin{split}
  Score &=  RoundTime \times RHP
\end{split}
\end{equation}
Here, $RHP$ shows the normalized sum of both players' lost HP in the current round as follows:
\begin{equation}
\begin{split}
  RHP &=  \frac{initialHP - 0.5(RHP_{P1} + RHP_{P2})}{initialHP},
\end{split}
\end{equation}

where $initialHP$ represents the initial HPs (Hit Points) of both Player1 (P1) and Player2 (P2). $RHP^{P1}$, $RHP^{P2}$ represents Remaining HP of P1 and P2, respectively. Equation 2 takes into account current remaining HP and starting HP of both players. If the total HP of P1 and P2 goes low, $RHP$ obtains a higher value. Besides, the game is tended to be more exciting if the game is close and less exciting vice versa, the value of $RHP$ would be higher for the close game which the total HP have a smaller amount than the game that one of the players has a large lead in remaining HP over the other. The term $Score$ adopted to the highlight evaluation of game commentator from ``Score Transition''\cite{ishii2019fighting} and ``Score Differential''\cite{hlcues}. 

\subsubsection{Action}
Four types of actions were highly evaluated as $RankAct$ in fightingICE highlights:  ``Special Skill,'' ``Strong Upper,'' ``Sliding Kick,'' and ``Shoot Strong Projectile Forward''\cite{ishii2019fighting}. Each of these four actions is considered to be exciting plays for the reason that having high visual effects. For the weight of specific actions, as proven in previous work, the empirical fixed value of $Rank$ used in the experiments are shown in Table \ref{tab:rankact}. In order to obtain a higher value of excitement in our work, we prioritize these actions using an equation as follows:

\begin{equation}
\begin{split}
      Action = 
        \begin{cases}
            {\frac{1}{2} + \frac{1}{2^{Rank}}} & \text{attack actions belongs to RankAct} \\
            \frac{1}{2} & \text{attack actions} \\
            0 & \text{otherwise}
        \end{cases}
\end{split}
\end{equation}

\subsubsection{Distance}
In this work, $Distance$ giving prominence to the close-distance fight between P1 and P2 and is defined as follows:

\begin{equation}
\begin{split}
  Distance &= 1 - \frac{\left| Xpos_{P1} - Xpos_{P2} \right|}{max(\left|Xpos_{P1} - Xpos_{P2}\right|))}
\end{split}
\end{equation}

where $Xpos$ is the $x$ coordinate of a player. $Distance$ obtains a higher value when P1 is positioned horizontal closer to P2, which we consider player has more effective actions, which can cause damage to its opponent player to perform. Note that because of highlight action generation by Ishii et. al\cite{ishii2019fighting} it determines actions for both players (P1 and P2). Their evaluation function is capable of engaging both P1 and P2 to get closer to the center. This could also let them positioned closer to $x$ coordinate of the center of the screen, which means closer to each other as well. As a result, gameplays that are tended to be exciting but positioned at the edge of the screen, are not existent in their work. However, in our mechanism, we did not determine action for both players, so such plays are unavoidable and also the one which needs to be highly evaluated.

\subsubsection{Highlight Formula for Dynamic Adjuster}
The cues $Score$, $Action$ and $Distance$ have been normalized and combined into $Highlight$ using the weighted sums $\omega_{s}$, $\omega_{a}$ and $\omega_{d}$ in equation 6 which was proven the most effective in extracting highlights by giving those actions higher value. The final Highlight Formula for Dynamic Adjuster is defined as follows:

\begin{equation}
\begin{split}
  Highlight = \omega_{s}Score + \omega_{a}Action + \omega_{d}Distance
\end{split}
\end{equation}

where $Score$, $Action$, and $Distance$ are normalized to the range of 0 to 1. And the proven weight settings\cite{ishii2019fighting} with best performance among other parameter settings in their work, is given as follows: 

\begin{equation}
\begin{split}
\omega_{s} = \omega_{a} = \omega_{d} = \frac{1}{3}
\end{split}
\end{equation}

\subsection{Commentary Text Generation for FightingICE}

Templates were built based on commentary scripts of \textit{King of the Fighters} competition tournament videos crawled from Youtube. Names of the skills and the characters were replaced for FightingICE. 104 professional commentary templates were made.

\subsection{Text-to-Speech}
Recent advances in generative models for TTS systems have enabled them to synthesize speeches almost as natural as human speakers~\cite{naturaltts}. Cloud speech synthesis platforms such as Google Cloud provide customization of synthesized voices through phonetic parameters such as volume, pitch, and rate. Such customization provides the potential of using TTS systems to build auditory commentary systems. In this study, we adopt a TTS system using the WaveNet model~\cite{oord2016wavenet} provided by Google Cloud to synthesize human-like voices for commentaries.  The synthetic voice for each commentary is ``en-US-Wavenet-D,'' a male voice speaking in the US accent. The range of each of the aforementioned two phonetic parameters -- pitch and volume -- used in the conducted experiment is given in Table~\ref{tab:ttsparameter}, while the default values are used for the other phonetic parameters. In this table, we empirically derived the ranges through our pilot studies and from previous work\footnote{https://cloud.google.com/text-to-speech/docs/reference/rest/v1/text/synthesize} using the same TTS system.


\begin{table}[tb] 
  \caption{Ranges of the Phonetic Parameters used in TTS} 
  \label{tab:ttsparameter}
  \hbox to\hsize{\hfil
  \begin{tabular}{c c c}
    \toprule
    Phonetic Parameter & Range & Default value\\
    \midrule
    Pitch &  [-6, 14] & 0\\
    Volume & [-6, 6] & 0\\
    \bottomrule
    \hline
  \end{tabular}\hfil}
\end{table}

\section{User Study}

We conducted a user study to investigate the preference of audiences towards five different designs for pitch and loudness control. Based on the value of \textit{Highlight} obtained from an AI by Ishii et al.\cite{ishii2019fighting}, we compare five different ways to adjust pitch and/or loudness of the game commentary (See Table 6); \textbf{Design 1}: no phonetic change, \textbf{Design 2}: increase loudness, \textbf{Design 3}: decrease loudness, \textbf{Design 4}: increase pitch, \textbf{Design 5}: decrease pitch. Five hypotheses were formulated under the context of FightingICE.
\begin{itemize}
    \item \textbf{H1:} The baseline (Design 1) is the least preferable. In other words, any design that changes phonetics based on \textit{Highlight} could positively the gameplay. 
    \item \textbf{H2:} During highlighted periods, increasing the loudness of the commentator (Design 2) entertain the gameplay. 
    \item \textbf{H3:} During highlighted periods, increasing the pitch of the commentator (Design 4) entertain the gameplay. 
    \item \textbf{H4:} Design 2 is better than Design 3. 
    \item \textbf{H5:} Design 4 is better than Design 5.
\end{itemize}








\subsection{Participants}
We conducted a user study using an online survey with 39 college students (33 males and 6 females, age $24.4 \pm 1.7$). They were asked to choose the most and the least preferable videos from a set of 5 shuffled videos. They were videos of the same gameplay, but with different commentaries based on 5 different designs above. 

\subsection{Experimental Enviorment}
\subsubsection{FightingICE}
The experiments were using conducted on recorded gameplay videos of FightingICE, which a fighting game platform for AI development and competition. The maximum time for each round of game in FightingICE is 60 seconds, and the game is rendered 60 frames per second. Besides, the initial HP for each character was equally set to 200. When the current game time of a round lasts 60 seconds or the HP of either character is reduced to 0, such round ends.

\subsubsection{TTS and Phonetic Variations Settings}
For comparing the effects of different phonetic aspects, The settings of five different designs were given in Table \ref{tab:combination}, where 1 represents setting the maximum value of that and 0 represents minimum except for the baseline Design 1 without changing any aspect, each of the designs only increases or decreases one aspect respectively.

\begin{table}[tb] 
  \caption{Phonetic variations used in experiment designs\\ (see value of \textit{Default} in Table \ref{tab:ttsparameter})} 
  \label{tab:combination}
  \hbox to\hsize{\hfil
  \begin{tabular}{c c c}
    \toprule
    Design & Volume & Pitch \\
    \midrule
    Design 1 & \textit{Default} & \textit{Default} \\
    Design 2 & $Highlight$ & \textit{Default} \\
    Design 3 & 1 - $Highlight$ & \textit{Default} \\
    Design 4 & \textit{Default} & $Highlight$ \\
    Design 5 & \textit{Default} & 1 - $Highlight$ \\
    \bottomrule
    \hline
  \end{tabular}\hfil}
\end{table}

\subsection{Results and Discussions}

\begin{table}[tb]
    \caption{Preferences of Video with p-value from Chi-square goodness of fit test.}
    \label{tab:comparison}
      \hbox to\hsize{\hfil
  \begin{tabular}{lrrrrrr}
    \toprule
    Design & 1 & 2 & 3 & 4 & 5 & $p$-value\\
    \midrule
    Best & 1 & 14 & 5 & 9 & 10 & 0.017 \\
    Worst & 13 & 1 & 5 & 10 & 10 & 0.020\\
    \bottomrule
    \hline
  \end{tabular}\hfil}
\end{table}


The result is shown in Table \ref{tab:comparison}. As expected, the majority of participants (13 out of 39) chose Design 1 as the least preferable; the standardized residual is 3.467, which is greater than 2, indicating that Design 1 is significantly differently prefered, thus \textbf{H1} holds true. \textbf{H2} was proven true as Design 2 was the most preferable (chosen by 14 out of 39 participants; the standardized residual is 2.220). \textbf{H4} holds true as while Design 2 significantly entertains participants, Design 3 was neither significantly chosen as the most or the least entertaining one. Unfortunately, Design 4 and Design 5 were neither chosen as the most or the least enter training (standardized residuals are in between -2 and 2), thus \textbf{H3} and \textbf{H5} were rejected. 

\subsubsection{chosen as the least entertaining} 
the reason ``plain commentary makes it boring,'' ``no emotional change on commentator'' of why Design 1 was chosen echoed our \textbf{H1}. the reasons for choosing Design 2 or Design 3 (related to our volume adjustment), ``the volume commentary of flickered high and low,'' ``sometimes could not able to hear commentary clearly'' point out the same reason of the depression from the drastic changes in volume. ``The change of pitch somehow changes'' was answered as the reason of choosing Design 4 or Design 5 (related to our pitch adjustment)

\subsubsection{chosen as the most entertaining} 
As we expected, for the designs adjusting on volume, participants chose Design 2 for ``The volume varies as the play goes,'' ``changes on volume when there is a sticky play'' reflected that our methods could make the video more entertaining. To our surprise, the response of ``The low pitch'' was also given, which shows some audiences might prefer the commentator with the fixed pitch. Referring to the finding on different pitch usage of English on people who are using different first language\cite{difflang,l1l2}, or different genders\cite{langgenderdiff}, it is also possible that some people may find such pitch variation or the variation set in our study is strange and prefer a fixed pitch or other variations from person to person.

For the designs adjusting on the pitch, although ``the change of pitch suit the gameplay'' and ``can feel the passion from the change of pitch'' were reported. Also,  comparing to Design 1, both Design 4 and Design 5 have a better result, which means there are positive effects when the pitch adjustment occurs. However, the result of pitch adjustment could not be able to support, which could perform better because there is no difference between Design 4 and Design 5. Besides, referring to the reason mentioned above that there are also participants who prefer the low, consistent pitch. In this section, we can conclude that the effect of adjusting pitch, adopting the excitement of the gameplay, may vary from person to person by genders or the language they use.

\section{Conclusion and Future Work}

To build an autonomous commentator for game videos and live streaming, this paper presented the first study of using the Text-to-Speech (TTS) mechanism to build the AI commentator and its correspondent method of getting the connection between highlight scenes of gameplay and phonetic variation by utilizing highlight cues. Our results from the survey in the conducted user study show that our method effectively affects the entertainment of gameplay video. For phonetic adjustment, in terms of volume variation, the results support that comparing among other Designs, increasing the commentator's volume when the gameplay is more entertaining, Design 2, is the most effective method that could be able to enhance the entertainment of the game videos. In terms of pitch variation, neither increase nor decrease has significantly improved entertainment. The results did not indicate which factors can be used to adjust the pitch effectively.
Our future work includes building a large scale dataset of commentary given by professional commentators that enable using neural networks to learn the highlight features instead of handcrafting them manually and investigating the difference in audience preferences based on gender, age, and nationality/culture. The proposed concept of integrating highlight cues for dynamic adjustment of commentary phonetic parameters can be applied to other games but requires a model for measuring the highlight of the game.


\end{document}